\documentclass[aps,pra,twocolumn,amsmath,amssymb,superscriptaddress]{revtex4-1}

\pdfoutput=1

\usepackage{graphicx}
\usepackage{dcolumn}
\usepackage[ansinew]{inputenc}
\usepackage{bm}
\usepackage{amsfonts}

\usepackage{color}

\begin{document}

\author{A.~Perrin}
 \affiliation{Vienna Center for Quantum Science and Technology, Atominstitut, TU Wien, 1020 Vienna, Austria}
 \affiliation{Wolfgang Pauli Institute, 1090 Vienna, Austria}
  \affiliation{Laboratoire de Physique des Lasers, Universit\'e Paris 13, CNRS, 99 Avenue J.-B. Cl\'ement, 93430 Villetaneuse, France}
\author{R.~B\"ucker}
 \affiliation{Vienna Center for Quantum Science and Technology, Atominstitut, TU Wien, 1020 Vienna, Austria} 
\author{S.~Manz}
 \affiliation{Vienna Center for Quantum Science and Technology, Atominstitut, TU Wien, 1020 Vienna, Austria} 
\author{T.~Betz}
 \affiliation{Vienna Center for Quantum Science and Technology, Atominstitut, TU Wien, 1020 Vienna, Austria} 
\author{C.~Koller}
 \affiliation{Vienna Center for Quantum Science and Technology, Atominstitut, TU Wien, 1020 Vienna, Austria} 
\author{T.~Plisson}
 \affiliation{Vienna Center for Quantum Science and Technology, Atominstitut, TU Wien, 1020 Vienna, Austria} 
 \affiliation{Laboratoire Charles Fabry de l'Institut d'Optique, Univ Paris Sud, CNRS, Campus Polytechnique RD128, 91127 Palaiseau, France}
\author{T.~Schumm}
 \affiliation{Vienna Center for Quantum Science and Technology, Atominstitut, TU Wien, 1020 Vienna, Austria}
\author{J.~Schmiedmayer}
  \affiliation{Vienna Center for Quantum Science and Technology, Atominstitut, TU Wien, 1020 Vienna, Austria}
\title{Hanbury Brown and Twiss correlations across the Bose-Einstein condensation threshold}

\date{\today}

\maketitle

\textbf{Hanbury Brown and Twiss (HBT) correlations, i.e. correlations in far-field intensity fluctuations, yield fundamental information on the quantum statistics of light sources, as highlighted after the discovery of photon bunching~\cite{HanburyBrown1956,Fano1961,glauber:63c}. Drawing on the analogy between photons and atoms, similar measurements have been performed for matter-wave sources, probing density fluctuations of expanding ultracold Bose gases~\cite{yasuda:96,foelling:05,schellekens:05,jeltes:07a,Hodgman2011}. Here we use two-point density correlations to study how coherence is gradually established when crossing the Bose-Einstein condensation (BEC) threshold. Our experiments reveal a persistent multimode character of the emerging matter-wave as seen in the non-trivial spatial shape of the correlation functions for all probed source geometries from nearly isotropic to quasi-one-dimensional (quasi-1D), and for all probed temperatures.  The qualitative features of our observations are captured by ideal Bose gas theory~\cite{gomes:06}, the quantitative differences illustrate the role of particle interactions.
}

HBT correlations can be related to a quantum interference effect reflecting the multimode nature of the source~\cite{Fano1961}. In analogy to thermal light sources, atom bunching in expanding thermal Bose gases has been observed~\cite{yasuda:96,foelling:05,schellekens:05,jeltes:07a,Hodgman2011}. Well above condensation threshold thermal Bose gases are sufficiently dilute so that atom-atom interactions are negligible and ideal gas theory provides an accurate description.

Extending the atom-photon analogy into the quantum degenerate regime, the absence of bunching has been observed in an out-coupled atom-laser~\cite{Ottl2005,Dall2011} and in an expanding BEC~\cite{schellekens:05,Hodgman2011}. This suggested a perfect coherence of these systems, similar to a monomode optical laser where interferences leading to HBT correlations are essentially absent, yielding only spatially and temporally uncorrelated poissonian shot noise~\cite{Arecchi1965}. 

Yet, studies of first order coherence properties of Bose gases near the BEC phase transition have reveiled the importance of thermal excitations, reducing the coherence also below threshold~\cite{bloch:00, donner:07}. Moreover, atom-atom interactions have been identified as a cause for a multimode nature in very elongated degenerate Bose gases. Originally predicted for the limit of weakly interacting degenerate 1D systems~\cite{Petrov2000}, similar behaviour is also present for very elongated 3D degenerate Bose gases~\cite{petrov:01,imambekov:09}. This effect has been demonstrated experimentally through measurements of density fluctuations~\cite{dettmer:01} or two-point correlations~\cite{Manz2010} for temperatures significantly below the BEC threshold.

In the following, we probe the second order correlation function $g_2$ of expanding Bose gases across the BEC phase transition, at zero and finite distances, to map out the gradual establishment of matter-wave coherence. Varying the source geometry, we explore the regime from \lq 3D\rq \  to  \lq quasi-1D\rq \  physics. Close to the BEC threshold, the fluctuations of the most populated quantum modes are significant and common theoretical models describing interacting Bose gases, as in~\cite{Petrov2000,petrov:01,imambekov:09}, fail. We therefore restrict our analysis to a qualitative comparison to ideal Bose gas theory predictions, allowing us to cover all investigated regimes.

\begin{figure}
 \centering\includegraphics{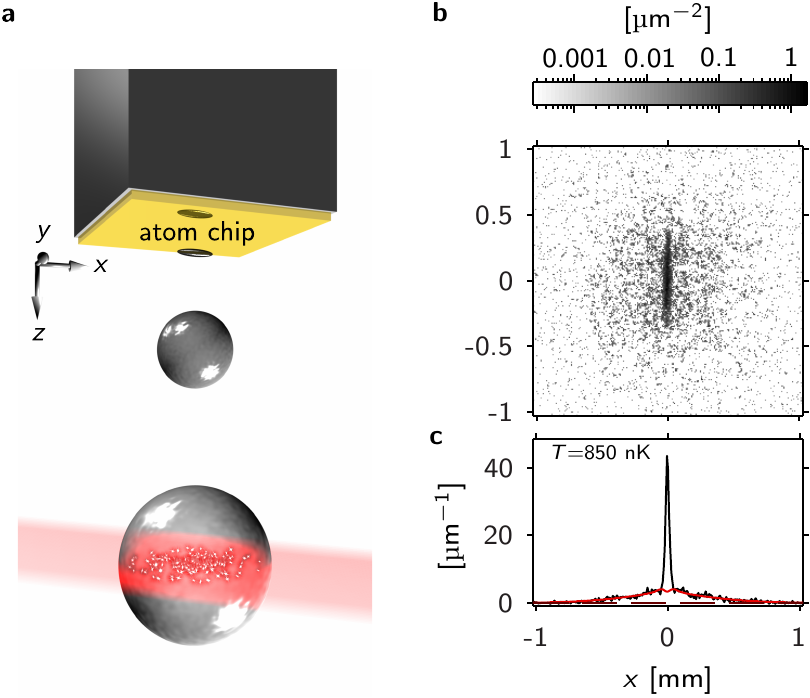}
 \caption{Fluorescence imaging of density correlations. \textbf{a}, Schematic of the experimental geometry. \textbf{b}, Example of the density distribution (lateral cut, log scale) of a Bose gas slightly below the Bose-Einstein condensation threshold after 46~ms expansion time. \textbf{c}, Profile of \textbf{b}, integrated along $y$. The red line indicates a fit to the density distribution of the thermal fraction, yielding the temperature $T$.}\label{fig:overview}
\end{figure}

\begin{figure*}
 \centering\includegraphics{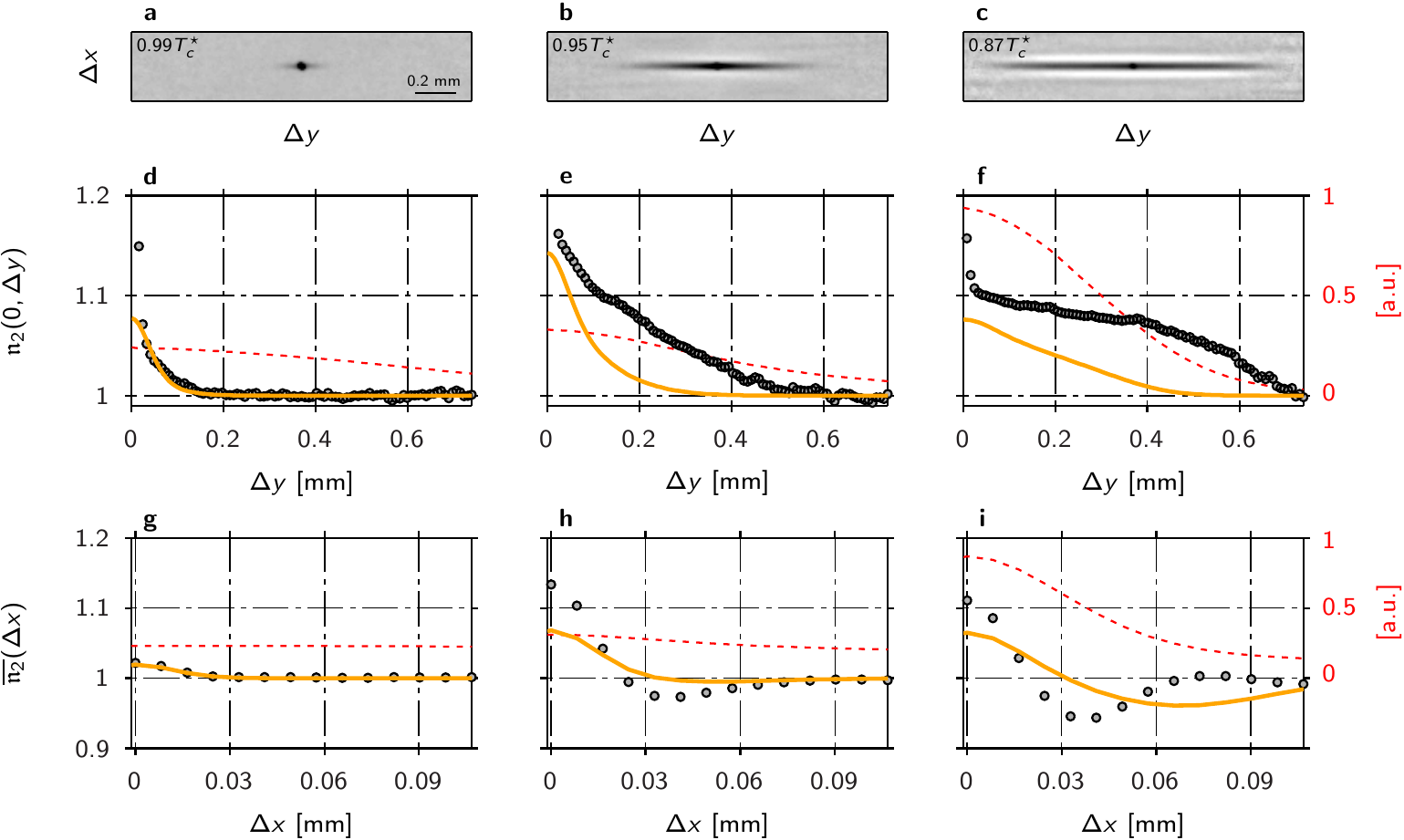}
 \caption{Density correlations results. \textbf{a},\textbf{b},\textbf{c}, Density correlation functions of expanding Bose gases $\mathfrak n_2$ (lateral cut in ($\Delta x$,$\Delta y$) plane) at $0.99\,T_c^\star$, $0.95\,T_c^\star$, and $0.85\,T_c^\star$, the aspect ratio $\lambda$ of the atomic source is 13.5. \textbf{d},\textbf{e},\textbf{f}, (circles) Radial cuts of \textbf{a},\textbf{b} and \textbf{c}. (dashed line) Radial cuts of the corresponding autocorrelation of the mean density profile. (solid line) Radial cuts of ideal Bose gas theory's predictions for the second order correlation function $\mathfrak g_2(0,\Delta y)$. \textbf{g},\textbf{h},\textbf{f}, (circles) Radially averaged axial cuts of \textbf{a},\textbf{b} and \textbf{c} over $160~\mu$m where the shot noise peak $\mathfrak s$ is excluded. The width of the exclusion region is $32~\mu$m. (dashed line) Radially averaged axial cuts of the corresponding autocorrelation of the mean density profile. (solid line) Radially averaged axial cuts of ideal Bose gas theory's predictions for the second order correlation function $\overline{\mathfrak g_2}(\Delta x)$.}\label{fig:bose_comparison}
\end{figure*}

The function $g_2(\mathbf{r},\mathbf{r'})$ measures the probability of joint detection of two particles at positions $\mathbf{r}$ and $\mathbf{r'}$ and hence relates to the density fluctuations of the system and their spatial correlations. For an ideal Bose gas above the BEC threshold $g_2(0)=2$, highlighting an excess of density fluctuations (bunching). For finite distances, $g_2(\mathbf{r},\mathbf{r'})$ decays to unity on the length scale of the temperature-dependent coherence length.  For a true monomode source $g_2(\mathbf{r},\mathbf{r'})=1$, demonstrating perfect coherence. In general the shape of $g_2(\mathbf{r},\mathbf{r'})$ reflects the interplay between the occupied modes of the system through the spatial scales of density fluctuations.

To obtain the second order correlation function of a Bose gas after its release from the trap, we use our novel fluorescence imaging~\cite{Bucker2009}.  Its high spatial resolution, single-atom sensitivity and exquisite signal-to-noise ratio enables us to probe the $g_2$ function with an accuracy at the percent level. We record a thin slice of the atomic density in the horizontal $x$-$y$ plane at the central part of the gas after 46~ms expansion (see Fig.~\ref{fig:overview}\textbf{a}-\textbf{b}). The thickness of the slice is set to $225~\mu$m by adjusting the duration of the excitation pulse to $500~\mu$s.

\begin{figure*}
 \centering\includegraphics{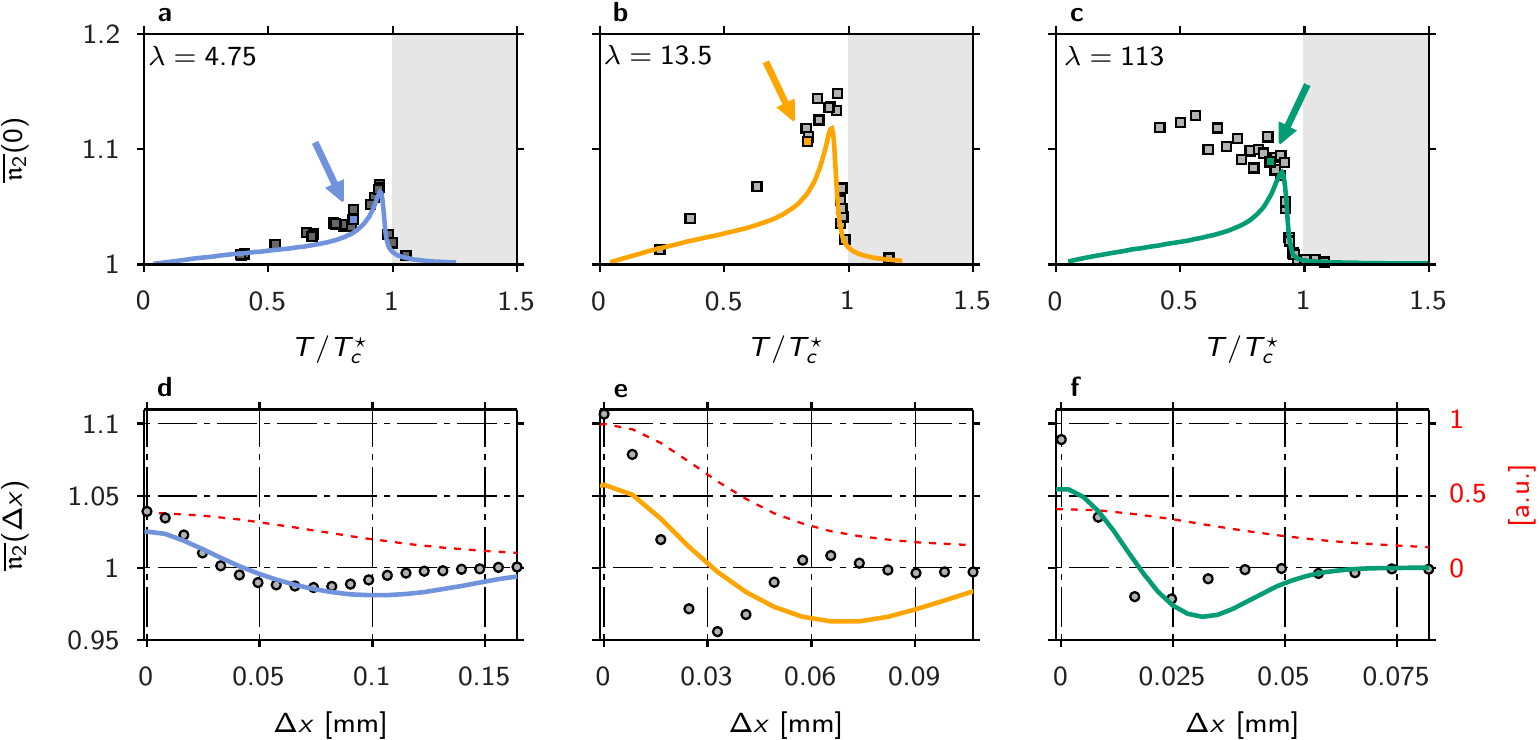}
 \caption{Density correlations comparison. \textbf{a},\textbf{b},\textbf{c}, Peak height of the second order correlation function of expanding Bose gases for various ratios $T/T_c^\star$ and three different trap aspect ratios $\lambda=4.75$ (light blue), 13.5 (orange) and 113 (green). It is defined as  $\overline{\mathfrak n_2}(0)$ (see caption Fig.~\ref{fig:bose_comparison}). The solid lines indicate the predictions obtained from ideal Bose gas theory. The light gray area display the range where the source is fully thermal. Note that the total atom number $N$ is not held constant between the data points, however this is taken into account in the value of $T_c^\star$. \textbf{d},\textbf{e},\textbf{f}, (circles) Radially averaged axial cuts of the density correlation function $\mathfrak n_2$ (see caption Fig.~\ref{fig:bose_comparison}) which parameters correspond to the colored square displayed respectively in \textbf{a},\textbf{b} and \textbf{c}. (dashed line) Radially averaged axial cuts of the corresponding autocorrelation of the mean density profile. (solid line) Radially averaged axial cuts of ideal Bose gas theory's predictions for the second order correlation function $\overline{\mathfrak g_2}(\Delta x)$.}\label{fig:g2_peak_height}
\end{figure*}

Our matter-wave source is a $^{87}$Rb gas of adjustable temperature prepared in the $|F,m_F\rangle=|1,-1\rangle$ state in an atom chip magnetic trap~\cite{Folman2002}. The chip design allows us to control the parameters of the magnetic trap and hence the shape of the source over a large range of aspect ratios $\lambda=\omega_{y,z}/\omega_x$ from 4 to 120, where $\omega_x$ and $\omega_{y,z}$ are the axial and radial trap frequencies respectively~\cite{trinker:08}. 

Modifying the aspect ratio of the trap we explore different regimes of Bose gases (see supplementary information). For the most isotropic traps ($\lambda=4.75,13.5$), gases significantly below the BEC threshold can be considered in the Thomas-Fermi regime with a chemical potential $\mu$ of the order of 3 to 4 $\hbar\omega_{y,z}$. 
In this case the expansion starts with a hydrodynamic phase. For the most elongated trap ($\lambda=113$), gases with temperatures below the degeneracy threshold have $\mu$ slightly smaller than $\hbar\omega_{y,z}$, entering the weakly interacting quasi-1D regime~\cite{gerbier:04c}. In this case the expansion of the system is essentially collisionless~\cite{imambekov:09}.

We obtain the temperature of the system $T$ and deduce the number of atoms in the thermal fraction of the gas $N_{\rm th}$ by fitting the thermal tails of the axial ($x$ axis) density profiles using an analytical formula based on ideal Bose gas theory (see Fig.~\ref{fig:overview}\textbf{c}). By comparing the fit result to the total observed signal, this also yields the number of condensed atoms (if present) within the measured slice. A simple model for the expansion of the degenerate fraction of the gas yields $N_{\rm bec}$, the total number of condensed atoms. Finally, for each set of temperature $T$ and total atom number $N=N_{\rm th}+N_{\rm bec}$, we estimate the critical temperature of the system $T_c^\star$ through the relation $T_c^\star=\alpha^{1/3}T_c$, where $T_c(T,N)$ corresponds to ideal Bose gas theory predictions for the critical temperature. The experimentally obtained factor $\alpha$ depends exclusively on the source aspect ratio $\lambda$ (see Methods).

Averaging over typically one hundred experimental repetitions with identical experimental parameters, we calculate the mean autocorrelation of the density profiles and normalize it by the autocorrelation of the mean density profile. We obtain the normalized density correlation function $\mathfrak n_2$, which can be decomposed into the sum of two terms, $\mathfrak s$ and $\mathfrak g_2$. The term $\mathfrak g_2$ contains the desired information about two-particle correlations whereas $\mathfrak s$ is a contribution due to atomic shot noise (see Methods). 

Typical results of density correlation functions of expanding Bose gases from a moderately anisotropic source ($\lambda=13.5$) are shown in Fig.~\ref{fig:bose_comparison} (the corresponding graphs for a nearly isotropic trap ($\lambda=4.75$, Fig.~S1) and a quasi-1D gas ($\lambda=113$, Fig.~S2) can be found in the supplementary information). The atomic shot noise $\mathfrak s$ appears as a dark oval spot at the center of the function. Its eigen-axes are rotated by $45^\circ$ with respect to the axes of the atomic source $x,y$, corresponding to the direction of the imaging laser beams. The structure corresponding to $\mathfrak g_2$ is visible behind this central spot. Its anisotropy reflects the aspect ratio of the source gas.

For a thermal gas above the BEC threshold ($T>T_c^\star$) we observe bunching. In this region, where inter-particle interactions are weak, we find an almost perfect agreement between the experimental observations and the second order correlation function $\mathfrak g_2$ simulated within ideal Bose gas theory. As input parameters we take the thermodynamical properties obtained from the data used to compute $\mathfrak n_2$ and account for the imaging resolution~\cite{gomes:06}. 

Below the critical temperature ($T<T_c^\star$) the main experimental observations valid for all explored aspect ratios (see Fig.~\ref{fig:bose_comparison} and Figs~S1 and S2 in the supplementary information) are:

(1) \emph{The establishment of coherence along the radial direction}: The RMS width of the $\mathfrak g_2$ peak along the $y$ axis rapidly grows when crossing the critical temperature $T_c^\star$ until it saturates due the finite radial size of the system (see Fig.~\ref{fig:bose_comparison}\textbf{d}-\textbf{f}). This can be seen as an experimental observation of the establishment of radial coherence~\cite{donner:07}, where the shape of $\mathfrak n_2$ changes from a decreasing exponential (Fig.~\ref{fig:bose_comparison}\textbf{d}-\textbf{e}) to a profile set by the spatial shape of the condensed cloud after expansion (Fig.~\ref{fig:bose_comparison}\textbf{f}). The RMS width of the $\mathfrak g_2$ peak is related to the radial coherence length in the trapped system through the hydrodynamic expansion of the gas, which imposes a scaling factor on the radial size of the cloud (see Methods).

(2) \emph{$\mathfrak g_2(0) > 1$ and the appearance of a dip below unity at finite distances along the axial $\mathfrak g_2$}:
Due to the anisotropy of the trap, thermal excitations of the system will typically be spread over more modes axially ($x$ axis) than radially. Below but close to the BEC threshold, few of the lowest lying axial modes will be macroscopically occupied and the shape of the second order correlation function $\mathfrak g_2(\Delta x>0)$ can then be interpreted as a result of the interference of all contributing modes. Hence the time-of-flight expansion implements a very sensitive heterodyne detection of weakly occupied modes of the matter-wave source.
The dip below unity at finite distance $\Delta x$ along the $\mathfrak g_2$ axial profiles is a direct consequence of this interference (see Fig.~\ref{fig:bose_comparison}\textbf{h}-\textbf{i}).  This observation is similar to observations reported in recent experiments and theoretical work on weakly interacting quasi-1D Bose gases~\cite{imambekov:09,Manz2010}  (for a comparison see Fig.~\ref{fig:g2_peak_height}\textbf{d}-\textbf{f}). Most interestingly we find this behaviour  - generally associated with quasi-1D physics - also for the most isotropic trap probed (aspect ratio $\lambda=4.75$).

We would like to point out that (2) corrects the widely established image of a perfectly \lq flat\rq \ correlation function for a Bose gas immediately below the BEC threshold~\cite{schellekens:05,Ottl2005,Dall2011,Hodgman2011}. Such behaviour highlights the influence of the non-ground-state modes (thermal depletion) of the system, whose population saturates below the BEC threshold. Working with small atom numbers and our highly sensitive detector allow us to accurately probe the transition regime and the graduate saturation of excitations when crossing the BEC threshold.

To give an overview we display the behaviour of $\overline{\mathfrak n_2}(0)$, set by $\mathfrak g_2(0)$, for all three measured aspect ratios and temperatures in a wide range of different ratios $T/T_c^\star$ in Fig.~\ref{fig:g2_peak_height}\textbf{a}-\textbf{c}.  Most strikingly we find $\mathfrak g_2(0) > 1$ for \emph{all} temperatures and aspect ratios, which indicates a persistent multimode nature even for a  three-dimensional BEC far below the condensation threshold. Lowering the temperature we observe a slow decrease of the amplitude of HBT correlations, which illustrates the gradual reduction of the thermal depletion of the system.

Ideal Bose gas theory reproduces the basic physics and the qualitative features of the experimental observation for moderate aspect ratios as discussed above also below the critical temperature. We attribute the remaining quantitative deviations between this description and our observations to particle interactions (see also Fig.~\ref{fig:g2_peak_height}\textbf{d}-\textbf{f} and Figs~S1 and S2 in the supplementary information).

For highly anisotropic sources with a large aspect ratio ($\lambda=113$) close to a quasi-1D system ($\mu<\hbar \omega$), we observe no reduction of $\mathfrak g_2(0)$ with temperature (see Fig.~\ref{fig:g2_peak_height}\textbf{c} and Fig.~S2 in the supplementary information). This behaviour is expected for very elongated systems where low lying axial excitations of the system remain macroscopically occupied also significantly below the degeneracy temperature~\cite{Petrov2000,petrov:01,imambekov:09,Manz2010}. It is based on the dominant influence of particle interactions, hence ideal Bose gas theory fails to describe even the qualitative behaviour at $T<T_c^\star$.

Our experimental findings call for a more complete theoretical description of interacting Bose gases at the threshold to Bose-Einstein condensation, where fluctuations of competing modes are significant. With such a theory at hand, measurements of density correlations will allow a detailed quantitative characterization of the mode occupation of the source. For equilibrium systems well below the BEC threshold this can be used for precision thermometry, where methods based on the observation of a thermal background fail (as demonstrated for deeply degenerate 1D systems~\cite{imambekov:09,Manz2010}). For ultralow temperatures well below the chemical potential, this would allow to probe quantum fluctuations, e.g. quantum depletion, of 3D or 1D Bose gases.  Our studies can directly be extended to non-equilibrium systems, where fundamental questions on equilibration, thermalization and integrability arise~\cite{mazets:08, Rigol2008,mazets:10}.

%

{\small
\section*{Methods}
\subsection*{Experimental Setup}

We prepare ultracold gases of a few $10^4$  $^{87}$Rb atoms in the $|F,m_F\rangle=|1,-1\rangle$ state confined in one out of 3 different Ioffe-Pritchard type magnetic trap obtained on top of a multilayer atom chip~\cite{trinker:08}. The 3 harmonic trap parameters are  $\omega_x=2\pi\times20$~Hz axially and $\omega_{y,z}=2\pi\times2260$~Hz radially  ($\lambda=113$), $\omega_x=2\pi\times160$~Hz  and $\omega_{y,z}=2\pi\times2180$~Hz ($\lambda=13.5$) and $\omega_x=2\pi\times320$~Hz and $\omega_{y,z}=2\pi\times1520$~Hz ($\lambda=4.75$) respectively.

We use forced radio-frequency (RF) evaporation to cool the gas close to or below the Bose-Einstein condensation threshold. As a direct consequence, colder gases contain less atoms and their corresponding critical temperature $T_c^\star=\alpha^{1/3}T_c$ decreases as well. Here $T_c=\hbar\overline{\omega}k_B^{-1}\left(N/\zeta(3)\right)^{1/3}$ where $N$ is the total number of atoms, $\overline{\omega}=\left(\omega_{x}\omega_{y}\omega_{z}\right)^{1/3}$, $\hbar$ and $k_B$ are Planck and Bolzmann constants respectively and $\zeta(x)$ is the Riemann zeta function. To ensure thermal equilibrium of the gas we keep a constant RF knife for 200~ms at the end of the evaporative cooling ramp.


\subsection*{Comparison with ideal Bose gas theory}

The constant $\alpha$ introduced in the main text is adjusted such that cooling the gas to $T_c^\star=\alpha^{1/3} T_c$ coincides with the experimental observation of the appearance of a degenerate fraction of atoms ($N_{\rm bec}\geq 0$). 
The value of $\alpha$ is slightly smaller than unity for the moderate trap aspect ratios ($\lambda=4.75,13.5$), which qualitatively agrees with theoretical studies of the shift of the critical temperature of interacting Bose gases~\cite{davis:06}. 
For the most elongated trap ($\lambda=113$), $\alpha$ is of the order of 3, which qualitatively agrees with predictions for the degeneracy temperature for 1D systems~\cite{kheruntsyan:03}.

\subsection*{Fit of the axial profile of the thermal fraction}

Using an analytic expression for the density flux $I$  of a freely expanding ideal Bose gas (see~\cite{gomes:06}, last formulas on page 3 and 5) and assuming an elliptical Gaussian shape for the excitation beams, hence neglecting any saturation effect, we deduce a formula describing the axial profile of density "slices" of the thermal fraction of expanding Bose gases
$$\rho_{\rm ls}(x)=\frac{1}{2\pi w}\int_{-\Delta t}^{\Delta t}dt\int dy dz \ \exp\left[-\frac{z^2}{2w^2}\right] I(x,y,z-z_0;t_0+t)$$
that we can use to fit our experimental results. Here $t_0$ is the expansion time, $z_0$ the position of the centre of the excitation laser beam profiles, $2\Delta t$ the exposure time and $w$ the RMS width of the excitation beams. For the data presented in this work $t_0=46$~ms, $2\Delta t=500~\mu$s and $w=10~\mu$m~\cite{Bucker2009}. This model depends on two parameters, the temperature of the gas $T$ and its fugacity, which fixes the total number of atoms in the thermal fraction of the gas $N_{\rm th}$. To avoid the divergence of the population of the ground state of the system when the fugacity reaches its maximum, only the excited states of the ideal Bose gas are considered for the calculation of $\rho_{\rm ls}(x)$.

\subsection*{Estimation of $N_{\rm bec}$}


For traps with moderate aspect ratios ($\lambda=4.75,13.5$), the condition $\mu\gg\hbar\omega_{x,y,z}$ is fulfilled quickly after passing the BEC threshold (see Fig.~S3 in the supplementary information) and the condensed part of the system can then be assumed to be in the Thomas Fermi regime. Scaling laws can then be used to describe the gas expansion~\cite{castin:96,kagan:96}. We find scaling factors of 6 (axial) and 625 (radial) for $\lambda=13.5$ and 28 (axial) and 430 (radial) for $\lambda=4.75$ respectively. With such model, deducing the number of condensed atoms $N_{\rm bec}$ from the number of condensed atoms counted in a density slice is straightforward.

For the most anisotropic trap ($\lambda=113$), we use an isotropic Gaussian ansatz for the radial density distribution of the degenerate fraction of the gas. Fitting the RMS width of the Gaussian along the $y$ axis, it is then straightforward to infer $N_{\rm bec}$.

\subsection*{Density correlation function}

The density correlation function is the sum of the two contributions $\mathfrak g_2$ and $\mathfrak s$. The first term $\mathfrak g_2$ can be expressed as $\mathfrak g_2=\overline{g_2}\ast o^2$, where $\overline{g_2}\left(\Delta\mathbf{r}\right)$ is an average of the second order correlation function $g_2(\mathbf{r},\mathbf{r}')$ over the spatially inhomogeneous density profile of the gas with $\Delta\mathbf{r}=\mathbf{r}-\mathbf{r}'$ kept fixed, $o^2$ is the effective two-particle point spread function (PSF) of the fluorescence imaging and $\ast$ is the convolution product. The function $o^2$ accounts for the diffusion of single atoms during the imaging process and optics aberrations, which blur the position of single atoms in the images~\cite{Bucker2009}. The other contribution to the density correlation function $\mathfrak s$ is due to the atomic shot noise and is proportional to the inverse of the mean density of the gas.

Here we slightly extend the analytic expression of the second order correlation function of an expanding ideal Bose gas derived in~~\cite{gomes:06} (see top right formula on page 6) to take into account the imaging scheme. Using a 2D isotropic Gaussian model for $o^2$ we obtain an analytic expression for $\mathfrak g_2$.

\section*{Acknowledgements}

We acknowledge support from the Wittgenstein prize, FWF projects F40, P21080-N16 and P22590-N16, the EU projects AQUTE and Marie Curie (FP7 GA no. 236702), the FWF doctoral program CoQuS (W 1210), the EUROQUASAR QuDeGPM Project and the FunMat research alliance. We wish to thank A. Aspect, D. Boiron, A. Gottlieb, I. Mazets, J. Vianna Gomes and C. I. Westbrook for stimulating discussions.

\section*{Author information}

Correspondence and requests for materials should be addressed to J. S. (jschmied@ati.ac.at).

\section*{Author contributions}

A. P., R. B., S. M. and T. S. collected the data presented in this letter. A. P. analyzed the data and developed the ideal Bose gas model. All authors contributed to the building of the experimental setup, the conceptual formulation of the physics, to the interpretation of the data and to writing the manuscript.

\section*{Competing financial interests}
The authors declare that they have no competing financial interests.
}

\section*{Supplementary Information}

\subsection*{Second order correlation results for $\lambda=4.75$ and $\lambda=113$}

\begin{figure*}
 \centering\includegraphics{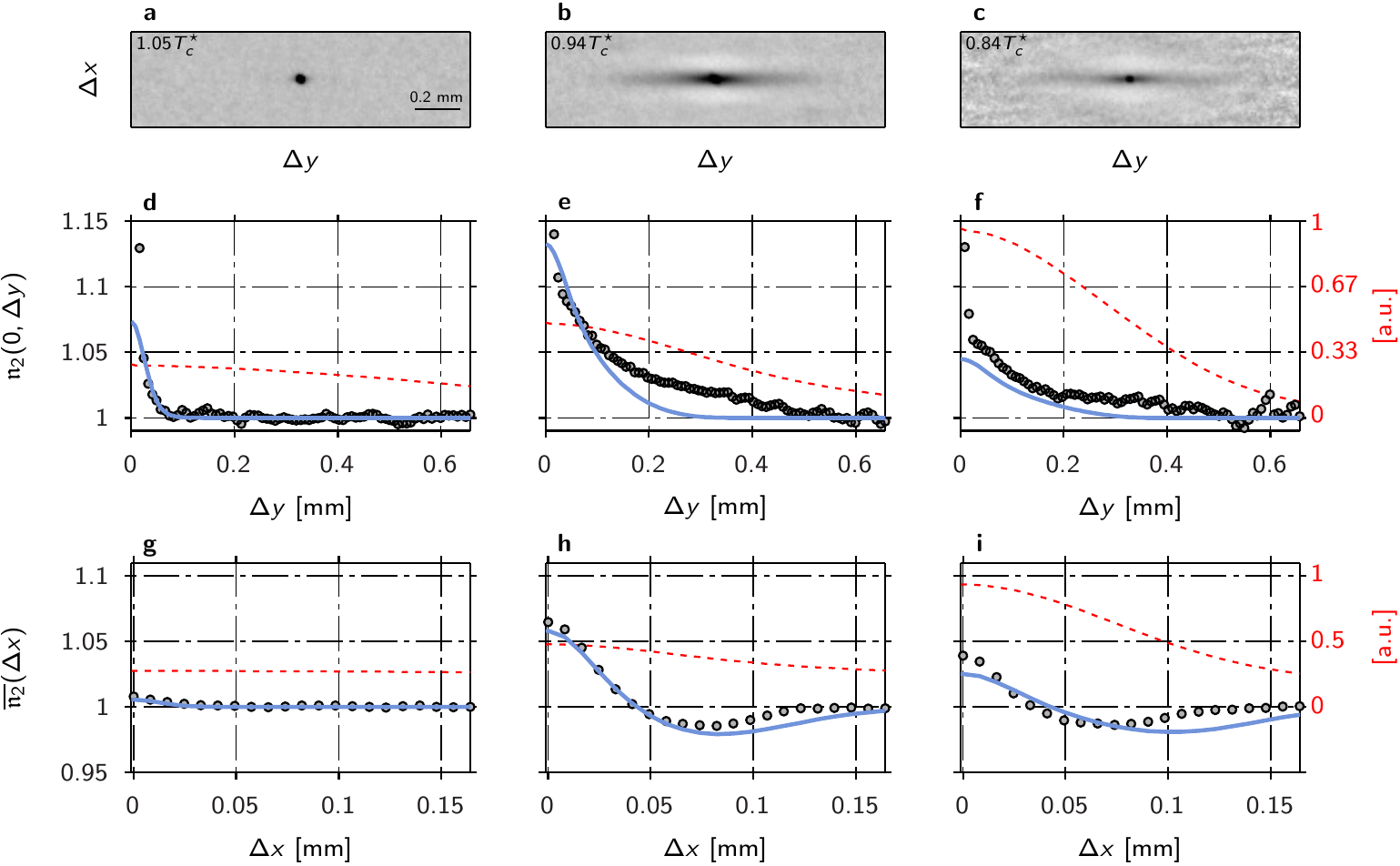}
 \caption{Density correlations results. \textbf{a},\textbf{b},\textbf{c}, Density correlation function of expanding Bose gases $\mathfrak n_2$ (lateral cut in ($\Delta x$,$\Delta y$) plane) at $1.05\,T_c^\star$, $0.94\,T_c^\star$, and $0.84\,T_c^\star$, the aspect ratio of the atomic source $\lambda$ is 4.75. \textbf{d},\textbf{e},\textbf{f}, (circles) Radial cuts of \textbf{a},\textbf{b} and \textbf{c}. (dashed line) Radial cuts of the corresponding autocorrelation of the mean density profile. (solid line) Radial cuts of ideal Bose gas theory's predictions for the second order correlation function $\mathfrak g_2(0,\Delta y)$. \textbf{g},\textbf{h},\textbf{f}, (circles) Radially averaged axial cuts of \textbf{a},\textbf{b} and \textbf{c} over $160~\mu$m where the shot noise peak $\mathfrak s$ is spatially excluded. The width of the exclusion region is $32~\mu$m. (dashed line) Radially averaged axial cuts of the corresponding autocorrelation of the mean density profile. (solid line) Radially averaged axial cuts of ideal Bose gas theory's predictions for the second order correlation function $\overline{\mathfrak g_2}(\Delta x)$.}\label{fig:bose_comparison2}
\end{figure*}

\begin{figure*}
 \centering\includegraphics{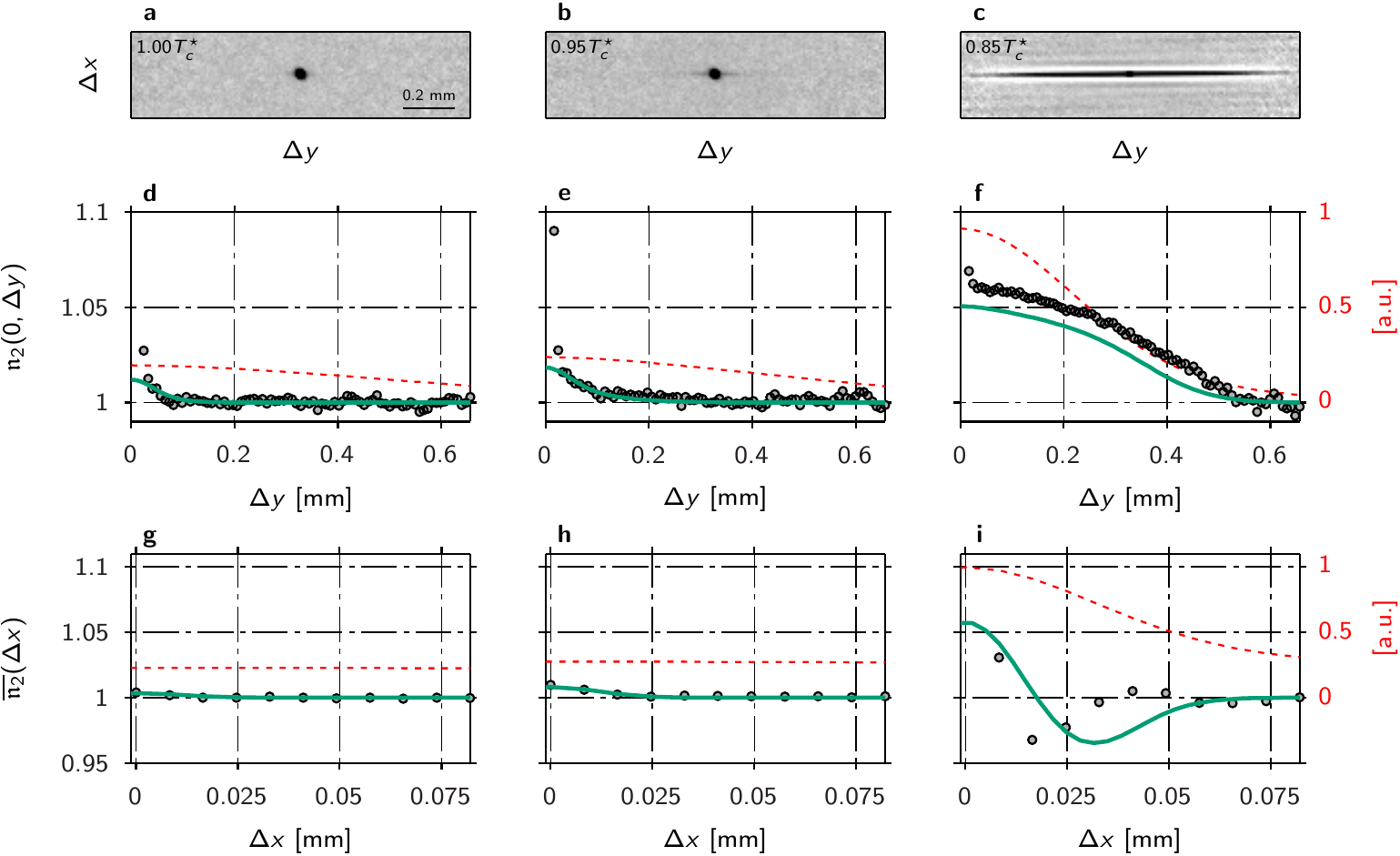}
 \caption{Density correlations results. \textbf{a},\textbf{b},\textbf{c}, Density correlation function of expanding Bose gases $\mathfrak n_2$ (lateral cut in ($\Delta x$,$\Delta y$) plane) at $1.00\,T_c^\star$, $0.95\,T_c^\star$, and $0.85\,T_c^\star$, the aspect ratio of the atomic source $\lambda$ is 113. \textbf{d},\textbf{e},\textbf{f}, (circles) Radial cuts of \textbf{a},\textbf{b} and \textbf{c}. (dashed line) Radial cuts of the corresponding autocorrelation of the mean density profile. (solid line) Radial cuts of ideal Bose gas theory's predictions for the second order correlation function $\mathfrak g_2(0,\Delta y)$. \textbf{g},\textbf{h},\textbf{f}, (circles) Radially averaged axial cuts of \textbf{a},\textbf{b} and \textbf{c} over $160~\mu$m where the shot noise peak $\mathfrak s$ is spatially excluded. The width of the exclusion region is $32~\mu$m. (dashed line) Radially averaged axial cuts of the corresponding autocorrelation of the mean density profile. (solid line) Radially averaged axial cuts of ideal Bose gas theory's predictions for the second order correlation function $\overline{\mathfrak g_2}(\Delta x)$.}\label{fig:bose_comparison3}
\end{figure*}

Here we present figures showing the same analysis data as for Fig.~\ref{fig:bose_comparison} in the main text but for different trap aspect ratios: $\lambda=4.75$ for Fig.~\ref{fig:bose_comparison2} and $\lambda=113$ for Fig.~\ref{fig:bose_comparison3}.

\clearpage

\subsection*{Thermodynamical properties of the data}

\begin{figure*}
 \centering\includegraphics{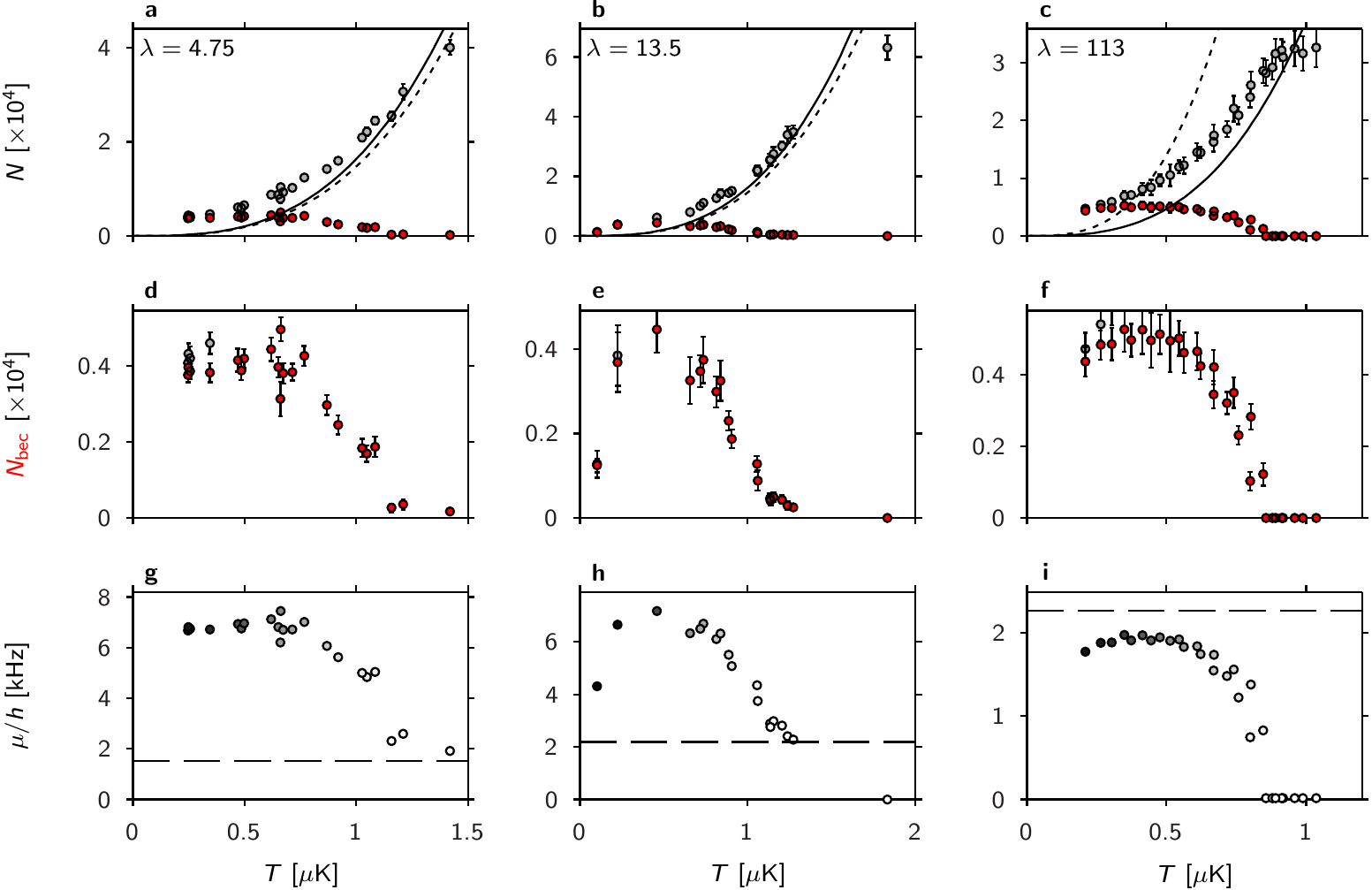}
 \caption{Thermodynamical properties of the data. \textbf{a},\textbf{b},\textbf{c} Scatter plot of the means of the derived total atom number $N$ (grey disks) and total condensed atom number $N_{\rm bec}$ (red disks) versus the temperature $T$, following the procedure described in the Methods section. The error bars give the standard deviation of $N$ and $N_{\rm bec}$ over individual samples.  The dashed line indicate the ideal Bose gas theory prediction for the critical atom number (see text). The solid line gives the same prediction corrected by the phenomenological factor $\alpha$. \textbf{d},\textbf{e},\textbf{f} Reproduction of \textbf{a},\textbf{b},\textbf{c} zooming on the behaviour of $N_{\rm bec}$. \textbf{g},\textbf{h},\textbf{i} Scatter plot of the estimate of the chemical potential of the gases. The color of each disk is related to the ratio $N_{\rm bec}/N$. The method used to obtain $\mu$ is described in the text. It is strictly valid only for the darkest disks. The dashed line indicates the value of the radial trapping frequency $\omega_{y,z}/2\pi$. }\label{fig:atom_number}
\end{figure*}

In Fig.~\ref{fig:atom_number} we show the thermodynamical properties of the different data used to compute the second order correlation functions. As explained in the methods section, the total atom number $N$ and the number of condensed atoms $N_{\rm bec}$ are obtained from the individual images using simple models for the expansion of the gas. According to ideal Bose gas theory, the condensation threshold is characterized at a given temperature by a critical atom number

$$ N_{\rm c}=\left(\frac{k_B T}{\hbar\overline{\omega}}\right)^3\zeta(3)$$
over which the excited states of the system are saturated and the atoms accumulate in the ground state of the trap. This prediction has to be rescaled using the phenomenological factor $\alpha$ to comply with our experimental observations, accounting for the effects of interactions and dimensionality.

We estimate our error on the temperature estimation of the mean density to be less than 10~\%. This gives an error on the estimation of the ratio $T/T_c^\star$ of the order of 10~\%. Nevertheless, close to $T_c^\star$, the measurement is much more accurate since the presence of a few condensed atoms is easily visible in individual images. This feature is used for the determination of $\alpha$.

To estimate $\mu$ in the case of the most isotropic trap ($\lambda=4.75,13.5$) we assume the system to be in the Thomas Fermi regime, in which case

$$ \mu_{\rm TF} = \frac{1}{2}\hbar\overline{\omega}\left[15\frac{N_{\rm bec}a}{\left(\hbar/m\overline{\omega}\right)^{1/2}}\right]^{2/5}$$
where $a=5.3$~nm is the s-wave scattering length and $\overline{\omega}=(\omega_{x}\omega_{y}\omega_z)^{1/3}$. Such assumption is only strictly valid when $ \mu_{\rm TF} \gg \omega_i$, with $i=x,y,z$, which is fulfilled in our case only for the coldest data.

In the case of the most elongated trap, we assume the system to be in the quasi-1D regime and estimate $\mu$ following Ref.~\cite{gerbier:04c}. Here again, the approach is only strictly valid for the coldest samples.

\clearpage

\subsection*{Normalisation of the fluctuations}

When averaging the fluorescence pictures in order to obtain the density correlation function $\mathfrak n_2$, we average clouds with slightly different total atom numbers. These fluctuations can be corrected by normalizing $\mathfrak n_2$ by the factor $1+{\rm Var}(S)/\overline{S}^2$, where $S$ is the total fluorescence signal in each picture, $\overline{S}$ its mean value for the considered clouds and ${\rm Var}(S)$ its variance. This factor differs from one by less than a percent for all the data presented here. No further normalization is necessary in order to obtain $\mathfrak n_2$ equal to unity at large distances $\Delta x$, $\Delta y$.

\subsection*{Imaging resolution}

While the resolution of the fluorescence imaging along the $z$ axis is fixed by the waist of the light sheet beams and the duration of the pulse, in the $x-y$ plane it is to a large extent determined by the diffusion of atoms during the detection process~\cite{Bucker2009}. Hence, each individual atom appears as a \emph{fluorescence pattern}, composed of typically 5 detected photons randomly distributed over a few adjacent pixels. The broadening of the atomic shot noise peak $\mathfrak s$ is given by the mean autocorrelation of the fluorescence patterns. Its elliptic shape reflects the increased diffusion of the atoms along the direction of the excitation beams. It is modelled by a 2D Gaussian with eigen-axes rotated by $45^\circ$ and fitted RMS widths $\sqrt{2}\times 5.5~\mu$m and $\sqrt{2}\times 6.2~\mu$m. However, $\mathfrak s$ is insensitive to the random offset of each fluorescence pattern centroid with respect to the actual position of the corresponding atom. This offset has to be included in a full description of the imaging resolution, yielding the two-particle PSF $o^2$.

Assuming a 2D isotropic Gaussian model for $o^2$, we can estimate its RMS width from the observed axial width of $\mathfrak g_2$ above the condensation threshold. The RMS width of $\overline{g_2}$, related to the axial coherence length of the gas~\cite{schellekens:05}, is expected to be much smaller than the observed RMS width of $\mathfrak g_2$. Thus, we assume the latter to originate only from the finite resolution (see Fig.~\ref{fig:bose_comparison}\textbf{g}). This, together with the finite resolution of the fluorescence imaging along the z axis, explains the reduction of $\mathfrak g_2(0)$ below 2 for systems close or above the BEC threshold~\cite{schellekens:05}. As the observed width stems from a two-particle correlation, it is sensitive to the actual resolution of the system, including the centroid offset. Finally, we obtain a value of $\sqrt{2}\times 8~\mu$m for the RMS width of $o^2$.

\end{document}